\documentclass[review]{elsarticle}

\usepackage{lineno,hyperref}
\usepackage{fullpage}
\usepackage{graphicx}
\usepackage{upref}
\usepackage{enumerate}
\usepackage{latexsym}
\usepackage{pdfpages}
\usepackage{color,graphics}
\usepackage{comment} 	
\usepackage{caption}
\usepackage{subcaption}
\usepackage{wrapfig}
\usepackage{braket}
\usepackage{amssymb}
\usepackage{amsmath}
\usepackage{amsthm}
\usepackage{mathrsfs}
\usepackage{mathtools}
\usepackage{algorithm,algorithmic,epsfig,graphicx}
\usepackage{amsfonts}
\usepackage{tikz} 
\usepackage{varioref}
\usepackage[ansinew]{inputenc}

\theoremstyle{plain}
\newtheorem*{theorem*}{}
\newtheorem{theorem}{Theorem}[section]
\theoremstyle{plain}
\newtheorem{lemma}{Lemma}[section]
\theoremstyle{plain}
\newtheorem{cor}{Corollary}[section]

\theoremstyle{plain}
\newtheorem{prop}{Proposition}[section]
\theoremstyle{plain}
\newtheorem{conj}{Conjecture}[section]
\theoremstyle{plain}

\theoremstyle{definition}
\newtheorem{defn}{Definition}[section]
\theoremstyle{definition}

\theoremstyle{remark}
\newtheorem{remark}{Remark}[section]
\theoremstyle{definition}

\newcommand{\E}{\mathbb{E}} 	
\DeclareMathOperator{\real}{\mathbb{R}}

\newcommand{\intg}{\mathbb{Z}}
\newcommand{\ratn}{\mathbb{Q}}

\newcommand{\poly}{\text{poly}}

\newcommand{\id}{\mathbb{I}}

\newcommand{\p}{\text{P}}
\newcommand{\NP}{\text{NP}}

\newcommand{\coNP}{\text{coNP}}

\newcommand{\DTIME}{\text{DTIME}}
\newcommand{\RTIME}{\text{RTIME}}
\newcommand{\AM}{\text{AM}}

\newcommand{\cL}{\mathcal{L}}
\newcommand{\SVP}{\textsf{SVP}}
\newcommand{\svp}{\textsf{SVP}}
\newcommand{\svpi}{\textsf{SVP}_{\infty}}
\newcommand{\CVP}{\textsf{CVP}}
\newcommand{\cvp}{\textsf{CVP}}
\newcommand{\cvpi}{\textsf{CVP}_{\infty}}
\newcommand{\gapsvp}{\textsf{GapSVP}}
\newcommand{\gapcvp}{\textsf{GapCVP}}

\newcommand{\vect}[1]{\mathbf{#1}}

\renewcommand{\vec}[1]{\overrightarrow{#1}}

\newcommand{\fld}{\mathcal{F}}
\newcommand{\assign}{\mathcal{A}}
\newcommand{\sis}{\textsf{SIS}}
\newcommand{\ncp}{\textsf{NCP}}
\newcommand{\lc}{\textsf{LC}}
\newcommand{\ssat}{\textsf{SSAT}}
\newcommand{\lhp}{\textsf{LHP}}
\newcommand{\sat}{\textsf{SAT}}
\newcommand{\ssati}{\textsf{SSAT}_{\infty}}
\newcommand{\pgc}{\textsf{PGC}}
\newcommand{\wt}{\text{wt}}

\newcommand{\eps}{\varepsilon}
\renewcommand{\epsilon}{\varepsilon}
\newcommand{\sound}{s}
\newcommand{\agreesound}{s_{agr}}
\newcommand{\listsound}{s_{list}}
\newcommand{\nzassign}{\mathcal{R}_{\neq0}}
\newcommand{\good}{\mathcal{G}}
\newcommand{\ngood}{\overline{\mathcal{G}}}

\modulolinenumbers[5]

\journal{Journal of Computer and System Sciences}

\begin{document}

\begin{frontmatter}

\title{The Projection Games Conjecture and the Hardness of Approximation of super-SAT and related problems}

\author{Priyanka Mukhopadhyay\fnref{mukhopadhyay.priyanka@gmail.com}\corref{mycorrespondingauthor}}
\address{Institute for Quantum Computing $\&$ Department of Combinatorics and Optimization\\ University of Waterloo  
\\200 University Avenue West, Waterloo, ON N2L 3G1}
\fntext[myfootnote]{Much of this work was done while the author was in Centre for Quantum Technologies, National University of Singapore.}
\cortext[mycorrespondingauthor]{Corresponding author}
\ead{mukhopadhyay.priyanka@gmail.com, p3mukhop@uwaterloo.ca}




\begin{abstract}
The Super-SAT or SSAT problem was introduced by Dinur et al.\cite{2003_DKRS,2002_D} to prove the 
 NP-hardness of approximation of two popular lattice problems - Shortest Vector Problem(SVP) and Closest Vector Problem(CVP). They conjectured that SSAT is NP-hard to approximate to within a factor of $n^c$ ($c>0$ is constant), where $n$ is the size of the SSAT instance. In this paper we prove this conjecture assuming the Projection Games Conjecture(PGC), given by Moshkovitz\cite{2012_M}. This implies hardness of approximation of SVP and CVP within polynomial factors, assuming PGC.
 We also reduce SSAT to the Nearest Codeword Problem(NCP) and Learning Halfspace Problem(LHP), as considered by Arora et al.\cite{1997_ABSS}. This proves that both these problems are NP-hard to approximate within a factor of $N^{c'/\log\log n}$($c'>0$ is constant) where $N$ is the size of the instances of the respective problems.
 Assuming PGC these problems are proved to be NP-hard to approximate within polynomial factors.
\end{abstract}


\end{frontmatter}

\section{Introduction}

\subsection{$\ssat$ and lattice problems}

The Super-$\sat$ or $\ssat$ problem was introduced by Dinur et al. \cite{2003_DKRS} to prove the hardness of approximation of two
popular lattice problems - the Shortest Vector Problem ($\SVP$) and the Closest Vector Problem ($\CVP$). An $n$-dimensional 
\emph{lattice} $\cL$ is the set of integral linear combinations of $n$ linearly independent vectors in $\real^d$, called the \emph{basis} of the 
lattice. The goal of $\SVP$ is to find a shortest non-zero lattice vector. Given a target vector $\vec{t}$, $\CVP$ aims at
finding a closest lattice vector to it.

Algorithms for these lattice problems are well-studied and have applications in factoring polynomials over rationals~\cite{1982_LLL},
integer programming~\cite{1983_L,2011_EHN}, cryptanalysis~\cite{2001_NS}, checking the solvability by radicals~\cite{1983_LM}, 
solving low-density subset-sum problems~\cite{1992_CJLOSS}, cryptography 
\cite{1996_A,2009_G,2009_R,2013_BLPRS,2014_BV,2017_DLLSSS}.

The $\ssat$ problem is the gap version of $\sat[F]$, which is defined as follows : An instance of $\sat[F]$ consists of a set of
constraints or Boolean functions, called \emph{tests}. The variables in each test take values from a finite set $F$ and each test 
has a set of satisfying assignments for its variables. The goal is to attach one assignment to each test such that consistency is
maintained i.e. each variable gets the same value in all the tests in which it appears. The cardinality of the set of satisfying assignments is bounded by a polynomial in the number of variables.

In $\ssat$, we attach integer weights to each assignment and call it a \emph{super-assignment}. It is \emph{consistent} if for each variable
the sum of weights on each value is the same in all the tests with this variable. Values which get non-zero net weight are said to
be simultaneously \emph{assigned} to the variable. If each variable gets at least one value assigned we call it a \emph{non-trivial}
super-assignment. If for at least one test there exists at least one non-zero weighted assignment then we call it
\emph{not-all-zero} super-assignment. This also gives rise to the notion of \emph{norm} 
of a super-assignment and accordingly two variants of $\ssat$ has been defined - the one in \cite{2003_DKRS} for $\ell_1$ norm and 
another by Dinur in \cite{2002_D} for $\ell_{\infty}$ norm. 
An instance is accepted (YES) if each variable gets a single value everywhere giving a consistent super-assignment of norm $1$. The rejection criteria is slightly different in the two variants. Roughly, an instance is rejected (NO) if every consistent super-assignment satisfying some conditions, has norm greater than $g$. If after minimizing the norm of such a consistent super-assignment we get a value between $1$ and $g$, then the instance may be accepted or rejected i.e. any outcome (YES/NO) is fine. A more detailed explanation of these concepts have been given in Section \ref{sec:ssat}. The following hardness results have been proved in \cite{2003_DKRS,2002_D}. Suppose $n$ is the size of the $\ssat$ instance, which is the encoding size of the number of variables, tests and satisfying assignments and this is polynomial in the number of variables (Section \ref{sec:ssat}). 

\textbf{$\ssat$ Theorem }\cite{2003_DKRS} :
 \emph{$\ssat$ is $\NP$-hard for $g=n^{c/\log\log n}$ for some constant $c>0$.}

\textbf{$\ssati$ Theorem }\cite{2002_D} :
 \emph{$\ssati$ is $\NP$-hard for $g=n^{c/\log\log n}$ for some constant $c>0$.}

An approximation factor preserving reduction from $\ssat$ to $\CVP_p$ (where distance is measured in $\ell_p$ norm), 
for $1\leq p<\infty$ was given in \cite{2003_DKRS} and a similar reduction from $\ssati$ to $\cvpi$ and $\svpi$
(where distance or length are measured in $\ell_{\infty}$ norm) was given in \cite{2002_D}. Thus the authors conjectured that the 
$\ssat$ problems are hard within a polynomial factor, which would imply $\NP$-hardness of the above mentioned lattice problems
within polynomial approximation factor.
\begin{conj}
 $\ssat$ is $\NP$-hard for $g=n^c$ for some constant $c>0$.
 \label{conj:ssat}
\end{conj}
\begin{conj}
 $\ssati$ is $\NP$-hard for $g=n^c$ for some constant $c>0$.
 \label{conj:ssati}
\end{conj}

\subsection{Label Cover ($\lc$) and Projection Games Conjecture ($\pgc$)}
\label{subsec:labelCover}

An instance of a Label Cover ($\lc$) problem (also referred to as \emph{Projection Games}) consists of (i) a bipartite graph 
$G=(A,B,E)$; (ii) finite alphabets $\Sigma_A,\Sigma_B$ from which each vertex of $A$ and $B$ (respectively) are assigned a label; 
(iii) constraints or projections $\pi_e:\Sigma_A\rightarrow\Sigma_B$ for each $e\in E$. Given a labeling or assignment to the 
vertices, $\varphi_A:A\rightarrow\Sigma_A$ and $\varphi_B:B\rightarrow\Sigma_B$, we say an edge $e=(a,b)$ is satisfied if the 
corresponding projection constraint holds, i.e. $\pi_e(\varphi_A(a))=\varphi_B(b)$.
In the optimization version of this problem the task is to find a labeling that maximizes the number of satisfied edges. 
The decision version of this problem is of interest to us and by Label Cover ($\lc$) we denote this problem of distinguishing 
between the (YES) case that all edges are satisfied and the (NO) case when at most $s$ (\emph{soundness error}) fraction of the edges
are satisfied. There are other variants of this problem (e.g. \cite{1997_ABSS}) but in this paper we work with this one.

 A \emph{PCP Theorem} gives the hardness of $\lc$ as follows [\cite{1998_AS, 1998_ALMSS, 1998_R}]: \\
  \emph{Given an input of size $N$ for $\lc$ with alphabet size $k$, it is $\NP$-hard to distinguish between the case where
 all edges can be satisfied and the case where at most $s$ fraction of the edges can be satisfied. ($s$ and $k$ may be
 functions of $N$.)}
\\
Equivalently it can be stated that there is a reduction from (exact) $\sat$ to $\lc$. Raz and Moshkovitz \cite{2010_MR} proved the following result.

\textbf{Theorem 11 in \cite{2010_MR}} (re-phrased):
\emph{
 There exists $c>0$ such that for every $s\geq 1/N^c$, $\sat$ on input of size $n$ can be reduced to $\lc$ of size $N$ for
 $N=n^{1+o(1)}\poly(1/s)$. The $\lc$ is over an alphabet of size exponential in $1/s$ and has soundness error $s$. 
 The reduction can be computed in time linear in the size and the alphabet size of the $\lc$ instance. The $\lc$ is on a
 bi-regular graph whose degrees are $\poly(1/s)$. }

PCPs which achieve an $\lc$ instance of size $N=n^{1+o(1)}\poly(1/s)$ are called \emph{almost-linear size PCP} because of the 
exponent of $n$. The soundness error $s$ is at least $1/N$. Assuming $\p\neq\NP$ it can be shown that the alphabet size is
at least $1/s$. Certain PCP constructions manage to have an alphabet size of $\poly(1/s)$ at the cost of some other
parameters \cite{1998_R}. Thus Moshkovitz \cite{2012_M} conjectured that a similar alphabet size may be achieved in the reduction in \cite{2010_MR}.

\begin{conj}[\textbf{Projection Games Conjecture ($\pgc$)} \cite{2012_M}]

There exists $c > 0$ such that for every $\sound \geq 1/N^c$, $\sat$ on input of size $n$ can be efficiently 
reduced to $\lc$ of size $N = n^{1+o(1)} \poly(1/\sound)$ over an alphabet of size $\poly(1/\sound)$ and has soundness error 
$\sound$.

 \label{conj:pgc}
\end{conj}
Here again $\sound$ is at least $1/N$.
\subsection{Related work}

The Label Cover ($\lc$) problem was introduced by Arora et al. \cite{1997_ABSS} but with a slightly different formulation than what
has been stated in this paper. Roughly, in their variant there is a ``cost'' attached to the labeling of each vertex. The
approximation factor is given by the ratio of this cost between the NO and YES case. The authors proved this variant of $\lc$ is
$\NP$-hard up to an approximation factor of $2^{\log^{0.5-\epsilon}n}$ where $\epsilon>0$ is some constant and $n$ is the size of 
$\lc$ instance, under the assumption that $\NP \nsubseteq \DTIME(n^{\poly(\log n)})$ . They gave an approximation factor 
preserving reduction from $\lc$ to a number of other problems like $\CVP$, $\svpi$, Nearest Codeword Problem ($\ncp$), Min-Unsatisfy
problem and Learning Halfspace Problem ($\lhp$). They also proved that the above problems were $\NP$-hard for constant approximation
factors by a reduction from Set Cover.

In 2012, Moshkovitz~\cite{2012_M} reduced $\lc$ (the variant stated in Section \ref{subsec:labelCover}) to Set Cover and proved that
the latter is $\NP$-hard to approximate within $(1-\alpha)\ln n$ ($n$ being the instance size) for arbitrarily small $\alpha>0$. 
She applied the Projection Games Conjecture to the reduction in \cite{1997_ABSS} and concluded polynomial approximation factors are
hard for $\CVP$. Here the conjecture has also been used to study the behavior of CSPs around their approximability threshold.

The first $\NP$ hardness result for $\CVP$ in all $\ell_p$ norms and $\SVP$ in the $\ell_{\infty}$ norm was given by van Emde Boas
\cite{1981_vE}. $\SVP$ was proven to be $\NP$-hard to approximate within a constant factor in \cite{1998_A,1998_CN,2001_M}.
Khot \cite{2005_K} and later Haviv and Regev \cite{2012_HR} improved the approximation factor to $2^{\log^{1-\epsilon}n}$
under the assumption that $\NP \nsubseteq \RTIME(n^{\poly(\log n)})$.
In \cite{2017_BGS} it has been shown that for almost all $p \ge 1$, $\CVP$ in the $\ell_p$ norm cannot be solved in 
$2^{n (1-\eps)}$ time under the Strong Exponential Time Hypothesis~\cite{1999_IP}. A similar hardness result has also been obtained 
for $\SVP$ \cite{2018_AsD}.

However, there are barriers for showing stronger inapproximability results. For example, a $\sqrt{n/O(\log n)}$ factor $\NP$-hardness result would imply $\coNP\subseteq\AM$
\cite{2000_GG}, a factor $\sqrt{n}$ $\NP$-hardness for $\SVP$ would imply $\NP=\coNP$ \cite{2005_AR}, and thus the polynomial 
hierarchy collapses in all these cases.

\subsection{Our results and techniques}

\subsubsection*{Hardness of $\ssat$ and $\ssati$}

In this paper we prove Conjecture \ref{conj:ssat} and \ref{conj:ssati}, assuming the Projection Games Conjecture ($\pgc$). 
Specifically we prove the following:
\begin{theorem}
 Assuming the PGC, $\ssat$ is $\NP$-hard for $g=n^c$ for some constant $c>0$.
 \label{thm:ssat}
\end{theorem}
\begin{theorem}
 Assuming the PGC, $\ssati$ is $\NP$-hard for $g=n^c$ for some constant $c>0$.
 \label{thm:ssati}
\end{theorem}

We give a reduction to both $\ssat$ and $\ssati$ from a variant of Label Cover, introduced by Moshkovitz in \cite{2012_M}, let us 
call it the list-Label Cover, which we explain very briefly here. 
Here we can assign each $A$-vertex a list of labels. Two vertices $a_1,a_2\in A$ \emph{agree} on a 
label $y\in\Sigma_B$ for a vertex $b\in B$ if there exists at least one label in their respective lists such that these map to $y$
under the respective edge constraint functions. That is, there exists $x_1\in\varphi_A(a_1)$ and $x_2\in\varphi_A(a_2)$ such that 
$\pi_{e_1}(x_1)=\pi_{e_2}(x_2)=y$, where $e_1=(a_1,b), e_2=(a_2,b)\in E$. For this variant of $\lc$ a notion of 
\emph{list agreement soundness} has been defined which gives the fraction of $B$-vertices on which the $A$-vertices totally disagree
(i.e. no two $A$-vertices agree on any one label). 
A more detailed explanation of these concepts have been given in Section \ref{subsec:LC}. 

In our construction, each $A$-vertex corresponds to a variable in the $\ssat$ instance and each $B$-vertex is a test.
By a result of Moshkovitz \cite{2012_M}, we can bound the $B$-degree (degree of each $B$-vertex) by some large enough constant prime
power. For simplicity and without much loss of generality, we assume that $\Sigma_A$ is in bijective correspondence to some finite
field and the variables take values from $\Sigma_A$. For each test $\psi_b$ (corresponding to some $b\in B$) consider a $y\in\Sigma_B$ 
such that it has at least one pre-image in each of $b$'s neighbor in $A$, consider the following tuples:
\begin{eqnarray}
 \mathcal{R}_y(\psi_b) = \{(x_1,\ldots,x_{D_B}) : x_j \in \pi_{e}^{-1}(y) \text{ where } e=(a_j,b) \text{ and } a_j 
 \text{ is the } j^{th} \text{ neighbor of } b \} \nonumber
\end{eqnarray}
Then the total set of satisfying assignments for $\psi_b$ is :
$ \quad
 \mathcal{R}(\psi_b) = \bigcup_{y \in \Sigma_B} \mathcal{R}_y(\psi_b)  \quad
$. Since we assume that the PGC (Conjecture \ref{conj:pgc}) is true, so $|\Sigma_A|, |\Sigma_B|$ and hence $|\mathcal{R}(\psi_b)|$ is bounded by some polynomial in the number of variables. 

Next, we argue that a YES case of $\lc$ maps to a YES case of $\ssat$. If there exists a labeling that satisfies all edges then we can construct an assignment such that each variable is assigned a single value in every test, yielding a consistent super-assignment of norm 1 (Lemma \ref{lem:complete}).

For the soundness proof we need to show that a NO instance of 
$\lc$ maps to a NO case of $\ssat$. Instead, we give a contrapositive argument.
In the $\ell_1$ norm (Lemma \ref{lem:sound1}) we show that if there exists a consistent non-trivial super-assignment of norm less than $g$ then there exists 
a labeling of the vertices such that a certain fraction of the $B$-vertices do not have total disagreement. We give a kind of rejection sampling procedure List-Construction, by which every variable (or A-vertex) is assigned a list including all assigned values and some non-assigned values from the list of non-zero satisfying assignments of the tests in which it appears (or B-neighbors). We ensure that the list size of each A-vertex is same. Then we prove that there exists a test or B-vertex $\psi_b$ on which at least two of its A-neighbors agree i.e. it has a non-zero weighted satisfying assignment $r_b=(x_1,\ldots,x_{D_B})$ such that at least two of the values in this tuple are in the list of their corresponding variables.

We show a similar result in the $\ell_{\infty}$ norm (Lemma \ref{lem:soundi2}) if there exists a consistent non-trivial (which is obviously not-all-zero) super-assignment. If a consistent super-assignment of norm at most $g$ is not-all-zero but not non-trivial i.e. some tests have all assignments with zero weight, then also such a labeling exists if some conditions hold. Here we must observe that if the norm of a not-all-zero consistent super-assignment is somewhere between $1$ and $g$ then any answer is acceptable. By this contrapositive argument we show that in some cases we get a YES reply from the oracle. So this is sufficient to prove
the soundness in the $\ell_{\infty}$ norm.

The $\ssat$ problem played a central role in proving the $n^{c/\log\log n}$ factor $\NP$-hardness of $\CVP$ and $\svpi$ \cite{2002_D, 2003_DKRS}. But not much work has been done in probing the complexity of this problem. To the best of our knowledge, we first relate this problem to other non-lattice problems. In \cite{2002_D, 2003_DKRS} the authors reduced a PCP instance to an $\ssat$ instance. The variables are embedded into a geometric domain and then this domain is recursively encoded by multiple new domains, adding new variables along the way. The construction relies on strong error correcting properties of low degree fucntions to decode any consistent low norm super-assignment into a satisfying assignment for the original set of tests. We are not going into the detail of this construction. Interested readers are encouraged to refer to the original papers \cite{2002_D, 2003_DKRS}. Our reduction is relatively much simpler because of the PGC. We reduce a variant of the $\lc$ problem to $\ssat$, as described very briefly before (more detail in Section \ref{sec:lc2ssat}). In \cite{2012_M} the author shows how the complexity of this variant of $\lc$ relates to the ``traditionally defined'' $\lc$ problem, and hence to $\sat$. This is sufficient to derive new complexity results for $\ssat$.

\subsubsection*{Hardness of lattice problems and some related problems}

Dinur et.al. \cite{2003_DKRS,2002_D} gave approximation factor preserving reduction from $\ssat$ and $\ssati$ to some lattice 
problems like $\CVP,\cvpi,\svpi$ and a related problem - Short Integer Solution ($\sis$). As a corollary of Theorem \ref{thm:ssat}
and \ref{thm:ssati} we prove these problems are $\NP$-hard to approximate within polynomial factors, provided Projection Games 
Conjecture holds (Corollary \ref{cor:sis_pgc},\ref{cor:cvp_pgc},\ref{cor:svp_pgc}). These problems have been defined explicitly in
Section \ref{subsec:others}.

\subsubsection*{Hardness of Learning Halfspace Problem ($\lhp$)}

We briefly describe the Learning Halfspace Problem ($\lhp$) in the presence of malicious errors, as defined in \cite{1997_ABSS}. 
It has been defined more precisely in Section \ref{subsec:others}. Roughly, the input to this problem is a set of points which are
labeled by ``+'' or ``-'', according to the side of a hyperplane they lie in a finite-dimensional space. Since this problem arises
in the context of training a perceptron (a learning model) \cite{2017_MP}, so we say that the input is given to a ``learner'', which has
to output a hypothesis (i.e. a hyperplane) that correctly classifies as many points as possible. The \emph{error} of an algorithm 
is the number of misclassifications by its hypothesis, and the \emph{noise} of the sample (set of points) is the minimum error
achievable by any algorithm. The \emph{failure ratio} of an algorithm is the ratio of its error to noise.

We take the formulation of $\lhp$ given in \cite{1997_ABSS} and give a reduction from $\ssat$, thus proving that approximating
the minimum failure ratio of $\lhp$ is $\NP$-hard upto a factor of $n^{c/\log \log n}$ (Theorem \ref{thm:lhp}). Assuming $\pgc$
this factor can be improved to $n^c$ for some constant $c>0$ (Corollary \ref{cor:lhp_pgc}).

\subsubsection*{Hardness of Nearest Codeword Problem ($\ncp$)}

The input to the Nearest Codeword Problem ($\ncp$) is the generator matrix of an error-correcting code of length $n$ over a $q$-ary
alphabet. Given a target vector the goal is to find a codeword that is nearest to the target vector in Hamming distance. More
detail description has been given in Section \ref{subsec:others}. Note that this problem is not exactly equivalent to $\CVP$.

The fact that this problem may be related to $\ssat$ was hinted in \cite{2003_DMS}. However, no proof was given there. We give
a reduction from $\ssat$ to $\ncp$ which proves that it is $\NP$-hard to approximate the latter within a factor $n^{c/\log\log n}$
(Theorem \ref{thm:ncp}), which can be improved to $n^c$ ($c>0$ is a constant) assuming the Projection Games Conjecture 
(Corollary \ref{cor:ncp_pgc}).

\subsection{Future directions}

One obvious direction would be to prove the hardness of $\ssat$ and $\ssati$ (Conjecture \ref{conj:ssat} and \ref{conj:ssati}) 
without assuming the Projection Games Conjecture. Alternately, one may prove the $\pgc$ (Conjecture \ref{conj:pgc}). Either of
these will imply improved $\NP$-hard approximation factors for the problems considered in this paper, without any other
assumptions. 

To the best of our knowledge, apart from this work, the $\ssat$ problem has only been studied to prove the hardness of approximation of lattice problems.
Relating this problem to other problems might give interesting hardness results as well as algorithms for them. It might even
throw some light on the complexity of $\ssat$ itself.

\subsection{Overview of this paper}

We give all necessary preliminary definitions and notations in Section \ref{sec:prelim}. The reduction from $\lc$ to $\ssat$ appears
in Section \ref{sec:lc2ssat}, while the reduction from $\ssat$ to other computational problems are in Section \ref{sec:app}.


\section{Preliminaries}
\label{sec:prelim}

\paragraph{Notations}
We write $\ln$ for natural logarithm and $\log$ for logarithm to the base $2$. $\real, \ratn, \intg$ denote the set of real 
numbers, rational numbers and integers respectively. $\mathbb{F}_p$ denotes a field of order $p$.
We denote variables by bold letters.
We denote arrays by letters (lower case letters for $1$-dimensional arrays or vectors) with overhead arrow, e.g. $\vec{v}^n$ and
$\vec{M}^{\ell\times m\times n}$ (or $\vec{M}_{\ell\times m\times n}$).
We may drop the dimension in the superscript (or subscript) whenever it is clear from the context.
The $i^{th}$ co-ordinate of $\vec{v}$ is denoted by $v_i$ or $(\vec{v})_i$. The $(i,j,k)^{th}$ entry of $\vec{M}$ is denoted by
$M_{ijk}$ or $\vec{M}[i,j,k]$.
Sometimes we represent a matrix ($2$-dimensional array) as a vector of column (vectors) 
(e.g. $\vec{M'}^{m\times n} = [\vec{m'}_1 \vec{m'}_2 \ldots \vec{m'}_n] $ where each $\vec{m'}_i$ is an $m-$length vector).

\subsection{Label Cover}
\label{subsec:LC}

\begin{defn}[\textbf{Label Cover ($\lc$)}]

An instance of $\lc$ consists of (i) a \emph{bipartite graph} $G=(A,B,E)$; (ii) finite \emph{alphabets} $\Sigma_A, \Sigma_B$, 
such that each vertex in $A$ and $B$ is assigned a \emph{label} from $\Sigma_A$ and $\Sigma_B$ respectively; 
(iii) a set $\Pi$ of constraints consisting of \emph{projections} $\pi_e:\Sigma_A \rightarrow \Sigma_B, \quad \forall e \in E$.
Given a labeling to the vertices $\varphi_A:A \rightarrow \Sigma_A$ and  $\varphi_B:B \rightarrow \Sigma_B$, an edge $e=(a,b)$ is
\emph{satisfied} if $\pi_e(\varphi_A(a)) = \varphi_B(b)$. (With a slight abuse of notation, we sometimes drop the labelings
and simply write $\pi_e(a)=b$.)

We work with the promise problem where in a YES instance there exists a labeling that satisfies all edges and in a NO instance, for all possible
labeling to the vertices at most $s$ fraction of the edges can be satisfied. Such an instance is said to have \emph{soundness error}
$\sound$. The goal is to distinguish between these two cases.
The \emph{size of the label cover} is $N=|A|+|B|+|E|$ and the \emph{size of the alphabet} is 
$\max\{ |\Sigma_A|,|\Sigma_B| \}$. 
\end{defn}

Feige \cite{1998_F} defined a variant of $\lc$ (using the structure obtained from parallel repetition) where the soundness is 
determined by the fraction of $B$-vertices that have at least
two neighbors from $A$ that \emph{agree} on a label for them. 
To be more precise, we define the following terms.

\begin{defn}[{\bf Total disagreement}]

Let $(G=(A,B,E),\Sigma_A,\Sigma_B,\Pi)$ be an $\lc$ instance. $\varphi_A:A \rightarrow \Sigma_A$ is a labeling to
the $A$-vertices. We say that the $A$-vertices \emph{totally disagree} on a vertex $b \in B$ if there are no two 
neighbors $a_1,a_2 \in A$ of $b$, for which 
$ \quad 
 \pi_{e_1}(\varphi_A(a_1)) = \pi_{e_2}(\varphi_A(a_2)), \quad
$
where $e_1=(a_1,b), \quad e_2=(a_2,b)$.
\end{defn}

\begin{defn}[{\bf Agreement soundness}]

Let $\mathcal{G}=(G=(A,B,E),\Sigma_A,\Sigma_B,\Pi)$ be an $\lc$ instance.
We say that $\mathcal{G}$ has \emph{agreement soundness error} $\agreesound$, if the following holds: 
In the YES instance, there exists a labeling $\varphi_A:A \rightarrow \Sigma_A$, $\varphi_B:B \rightarrow \Sigma_B$ that 
satisfies all edges and in the NO instance, for any labeling $\varphi_A:A \rightarrow \Sigma_A$, the $A$-vertices are
in total disagreement on at least $1-\agreesound$ fraction of the $b\in B$.
 
\end{defn}

Moshkovitz \cite{2012_M} considered a variant of $\lc$ where each vertex can be assigned a list of $\ell$ labels
and an agreement is interpreted as agreement on \emph{one} of the labels in the list. 
We define the following terms related to this variant, which we call the \emph{list-Label Cover} (list-$\lc$).

\begin{defn}[{\bf List total disagreement}]

Let $\mathcal{G}=(G=(A,B,E),\Sigma_A,\Sigma_B,\Pi)$ be an $\lc$ instance. Let $\ell \geq 1$ and
$\widehat{\varphi_A}:A \rightarrow \binom{\Sigma_A}{\ell}$ is a labeling that assigns each $A$-vertex $\ell$
alphabet symbols. We say that the $A$-vertices \emph{totally disagree} on a vertex $b\in B$ if there are no two 
neighbors $a_1,a_2 \in A$ of $b$, for which there exists $\sigma_1 \in \widehat{\varphi_A}(a_1), \quad 
\sigma_2 \in \widehat{\varphi_A}(a_2)$, such that 
$\quad	\pi_{e_1}(\sigma_1) = \pi_{e_2}(\sigma_2), \quad$
where $e_1=(a_1,b), \quad e_2=(a_2,b) \in E$.
 
\end{defn}

\begin{defn}[\textbf{List agreement soundness}]

Let $\mathcal{G}=(G=(A,B,E),\Sigma_A,\Sigma_B,\Pi)$ be an $\lc$ instance. We say that $\mathcal{G}$ has \emph{list agreement soundness error} $(\ell,\listsound)$, if the following holds: 
In the YES instance, there exists a labeling $\varphi_A:A \rightarrow \Sigma_A$, $\varphi_B:B \rightarrow \Sigma_B$ that 
satisfies all edges and in the NO instance, for any labeling 
$\widehat{\varphi_A}:A \rightarrow \binom{\Sigma_A}{\ell}$, the $A$-vertices are in total disagreement on at least $1-\listsound$ 
fraction of the $b\in B$.
 
\end{defn}

The following result relates agreement soundness and list agreement soundness. 

\begin{lemma}[\cite{2012_M}]

Let $\ell \geq 1$ and $ 0<\agreesound<1$. An $\lc$ with agreement soundness error $\agreesound$ has list agreement 
soundness error $(\ell,\agreesound \ell^2)$.
 \label{lemm:list_soundness}
\end{lemma}

\subsection{SuperSAT ($\ssat$)}
\label{sec:ssat}

An $\ssat$ instance has a set $\Psi=\{\psi_1,\ldots,\psi_{n'}\}$ of tests over variables $V=\{\vect{v}_1,\ldots,\vect{v}_m\}$ which take values from a field $\fld$, called \emph{range of the variables}. An \emph{assignment} $\assign_{\Psi}:V\rightarrow\fld$ maps each variable to a value in $\fld$. For convenience, we can think an assignment as a tuple of field values (for its variables) and it is \emph{satisfying} if these values evaluate to some required value for the test. Each test $\psi$ has a list $\mathcal{R}_{\psi}$ of satisfying assignments for its variables and we attach some ``weight'' to each such assignment. For any assignment $r$ we use $r|_{\vect{x}}$ to denote the value of variable $\vect{x}$ in this assignment.

\begin{defn}[\textbf{Super-assignment to tests}]

A \emph{super-assignment} is a function $S$ mapping each $\psi \in \Psi$ to a value from $\intg^{\mathcal{R}_{\psi}}$. Thus
$\vec{S(\psi)}$ is a vector of integer coefficients, one for each $r \in \mathcal{R}_{\psi}$.

\end{defn}
Denote $\vec{S(\psi)}[r]$ as the $r^{th}$ co-ordinate of $\vec{S(\psi)}$.
A \emph{natural super-assignment} assigns each $\psi \in \Psi$ a unit vector $\vec{e_r} \in \intg^{\mathcal{R}_{\psi}}$
with a $1$ in the $r^{th}$ co-ordinate  (i.e. $\vec{S(\psi)}[r]=1$ and $\vec{S(\psi)}[r'] = 0$ for all $r' \neq r$).
A super-assignment is \emph{not-all-zero} if there is at least one test $\psi \in \Psi$ for which $\vec{S(\psi)} \neq \vec{0}$.

\begin{defn}[\textbf{Projection}]

Given a super-assignment $S:\Psi \rightarrow \bigcup_{\psi} \intg^{\mathcal{R}_{\psi}} $, the \emph{projection} of 
$\vec{S(\psi)}$ on a variable $\vect{x}$ of $\psi, \quad \pi_{\vect{x}}(\vec{S(\psi)}) \in \intg^{\mathcal{F}}$, is defined as follows:
$$ \forall a\in\mathcal{F}: \quad \pi_{\vect{x}}(\vec{S(\psi)})[a] = \sum_{r\in \mathcal{R_{\psi}}:r|_{\vect{x}}=a } \vec{S(\psi)}[r]
$$ 
\end{defn}
The notion of projection facilitates the definition of consistency between tests.

\begin{defn}[\textbf{Consistency}]
A super-assignment $S$ to the tests in $\Psi$ is \emph{consistent} if the projections of two tests on each mutual variable are equal,
i.e. for every pair of tests $\psi_i$ and $\psi_j$ with a common variable $\vect{x}$,
$$ \pi_{\vect{x}}(\vec{S(\psi_i)}) = \pi_{\vect{x}}(\vec{S(\psi_j)})  $$
 
\end{defn}
$S$ is \emph{non-trivial} if for every variable $\vect{x} \in V$ there is at least one test $\psi \in \Psi$ that is 
\emph{not cancelled} on $\vect{x}$, i.e. $\pi_{\vect{x}}(\vec{S(\psi)}) \neq \vec{0}$.
For a variable $\vect{x}$ we think of all the values $a\in \mathcal{F}$ receiving non-zero coefficients in 
$\pi_{\vect{x}}(\vec{S(\psi)})$
(i.e. values for which $\pi_{\vect{x}}(\vec{S(\psi)})[a] \neq 0$) as being ``simultaneously'' \emph{assigned} to $\vect{x}$ by $\psi$.
The non-triviality requirement implies each variable must be assigned at least one value.

To formally define the $\ssat$ problem we have to introduce the notion of norm of a super-assignment. Here we note that in 
\cite{2003_DKRS} the problem was defined for $\ell_1$ norm (though the derived results work for all $\ell_p$ norm where $1\leq p < \infty$). Dinur \cite{2002_D} introduced a related problem called the $\ssati$, where some definitions like norm of a super-assignment were modified. 

\begin{defn}[{\bf Norm of a Super-Assignment}]

For the problem $\ssat$ the \emph{norm of a super-assignment} $S$ is the average norm of its individual assignments :
$ \quad\|S\| = \frac{1}{|\Psi|} \sum_{\psi\in \Psi} \|\vec{S(\psi)}\|_1 $.

We call $\|\vec{S(\psi)}\|_1=\sum_{r\in\mathcal{R}_{\psi}} |\vec{S(\psi)}[r]|$ as the \emph{norm of a test} $\psi$.

For the problem $\ssati$ the \emph{norm of a super-assignment} $S$ is defined as :
$\quad  \|S\|_{\infty} = \max_{\psi \in \Psi} \|\vec{S(\psi)}\|_1 $.
 
\end{defn}

We now formally define both the $g-\ssat$ and $g-\ssati$ problem. The parameter $g$ is an 
approximation factor for the norm of a super-assignment.
\begin{defn}[\textbf{$g-\ssat$ (and $g-\ssati$)}]

The input instance 
$$ \mathcal{I}=\langle \Psi=\{\psi_1,\ldots,\psi_{n}\}, V=\{\vect{v}_1,\ldots,\vect{v}_m\},\{\mathcal{R}_{\psi_1},
\ldots,\mathcal{R}_{\psi_{n}}\}  \rangle $$ consists of 
(i) a set $\Psi=\{ \psi_1,\ldots,\psi_{n} \}$ of \emph{tests} over a common set 
$V=\{ \vect{v}_1,\ldots,\vect{v}_m \}$ of \emph{variables} that take values in a field $\mathcal{F}$ ;
(ii) for each test $\psi \in \Psi$ a list $\mathcal{R}_{\psi}$ of \emph{satisfying assignments} to its variables, called the
\emph{range of the test} $\psi$. 
The size of an instance is the encoding size of the number of variables, tests and satisfying assignments. The parameters $m,|\mathcal{F}|$ and $|\mathcal{R}_{\psi}|$ are always bounded by some polynomial in $n$ and hence the size of an instance is also bounded by some polynomial in $n$.

This is a promise problem where in the YES instance there is a consistent natural super-assignment for $\Psi$. In the NO instance of $\ssat$, for every \emph{non-trivial} consistent super-assignment $S$ for $\Psi$, $\|S\| \ge g$. While in the 
 NO instance of $\ssati$, for every \emph{not-all-zero} consistent super-assignment $S$ for $\Psi$, $\|S\|_\infty \ge g$. 
 The goal is to distinguish between the YES and NO instance of the respective problems.
 
\end{defn}

\subsection{Lattice problems}
\label{subsec:lattice}

\begin{defn}
 A \emph{lattice} $\cL$ is a discrete additive subgroup of $\real^{d}$.
 Each lattice has a \emph{basis} $\vec{B} = [\vec{b}_1, \vec{b}_2, \ldots, \vec{b}_n]$ where $\vec{b}_i \in \real^{d}$ and
 $  \cL=\cL(\vec{B}) = \Big\{ \sum_{i=1}^n x_i\vec{b}_i : x_i \in \intg \quad \text{ for } \quad 1 \leq i \leq n\Big\} $. \\
 We call $n$ the \emph{rank} of $\cL$ and $d$ as the \emph{dimension}.
\end{defn}

We consider the following lattice problems.
From here on, in all the definitions $c \geq 1$ is some arbitrary approximation factor (usually specified as subscript), 
which can be a constant or a function of any parameter of the lattice (usually rank).
For exact versions of the problems (i.e. $c=1$) we drop the subscript.
Typically, we define length in terms of the $\ell_p$ norm for some $1 \leq p \leq \infty$. Thus
$
\|\vec{x}\|_p := (|x_1|^p + |x_2|^p + \cdots + |x_d|^p)^{1/p}
$
for finite $p$ and 
$
\|\vec{x}\|_\infty := \max |x_i|\; .
$

\begin{defn}[\textbf{Shortest Vector Problem ($\SVP_c^{(p)}$)}]

Given a lattice $\cL$ with rank $n$ the goal is to find a shortest non-zero vector in the lattice.

In the promise version, usually denoted as $\gapsvp_c^{(p)}$ the goal is to distinguish between the YES instance when 
$\exists \vec{v}\in\cL\setminus\{\vec{0}\}$ such that  $\|\vec{v}\|_p\leq r$ (for some positive real $r$ given as input) and the 
NO instance when all non-zero vectors in the lattice have length greater than $c\cdot r $.

\end{defn}

\begin{defn}[\textbf{Closest Vector Problem ($\CVP_c^{(p)}$)}]

Given a lattice $\cL$ with rank $n$ and a target vector $\vec{t} \in \real^d$ the goal is to find a closest lattice vector to 
$\vec{t}$.

In the promise version, usually denoted as $\gapcvp_c^{(p)}$, the goal is to distinguish between the YES instance when 
$\exists\vec{v} \in \cL$ such that $\|\vec{v}-\vec{t}\|_p \leq r$ (for some positive real $r$ given as input) and the 
NO instance when $\forall\vec{v}\in\cL, \quad\|\vec{v}-\vec{t}\|_p > c \cdot r$.
 
\end{defn}

In this paper, with a slight abuse of notation we denote both the optimization and the promise versions of the above problems by
the same notation, i.e. $\SVP$ and $\CVP$ respectively.
\subsection{Other computational problems}
\label{subsec:others}

\subsubsection*{Shortest Integer Solution} 

\begin{defn}[\textbf{Shortest Integer Solution ($\sis_c$)} \cite{2003_DKRS}]
Given (i) a matrix $\vec{B'}\in \intg^{m\times n}$; (ii) target vector $\vec{t} \in \intg^m$ such that 
$\vec{t} \in \{ \vec{B'x} : \vec{x} \in \intg^n \}$ and (iii) $d\in\intg$, the goal is to distinguish between the
YES instance when $\exists\vec{z} \in \intg^n$ such that $\vec{B'}\vec{z} = \vec{t}$ and $\|\vec{z}\|_p\leq d$ and the
NO instance when $\forall\vec{z}\in \intg^n$ where $\vec{B'}\vec{z}=\vec{t}$ we have $\|\vec{z}\|_p>c \cdot d$.

\end{defn}

\subsubsection*{Nearest Codeword Problem}

An \emph{error-correcting code} $\mathcal{A}$ of \emph{block length} $n$ over a $q$-ary alphabet $\Sigma$ ($=\mathbb{F}_q$) is a 
collection of strings (vectors) from $\Sigma^n$, called \emph{codewords}. A \emph{linear code} $\mathcal{A}$ is a linear subspace
of $\mathbb{F}_q^n$ over base field $\mathbb{F}_q$ and it can be compactly represented by a \emph{generator matrix} 
$\vec{A} \in \mathbb{F}_q^{m\times n}$ such that $\mathcal{A} = \{ \vec{Ax} :\vec{x} \in \mathbb{F}_q^n\}$.

For any $\vec{v} \in \Sigma^m$, the \emph{Hamming weight} of $\vec{v}$ is denoted by $\wt(\vec{v}) = | \{i : v_i \neq 0 \} |$. 
The \emph{Hamming distance} between two vectors $\vec{u},\vec{v} \in \Sigma^m$ is 
$\|\vec{u}-\vec{v}\|_H=\wt(\vec{u}-\vec{v})$.

\begin{defn}[\textbf{Nearest Codeword Problem($\ncp_c$)} \cite{1997_ABSS}]
Given (i) a matrix $\vec{A}$ over $\mathbb{F}_q^{m\times n}$; (ii) a target vector $\vec{t} \in \mathbb{F}_q^m$ and 
(iii) an integer $d$, the goal is to distinguish between the 
YES instance when $\exists\vec{z}\in \mathbb{F}_q^n$ such that $\|\vec{A}\vec{z}-\vec{t}\|_H \leq d$, and the
NO instance when $\forall\vec{z}\in \mathbb{F}_q^n, \quad \|\vec{A}\vec{z}-\vec{t}\|_H > c\cdot d$.
\end{defn}

\subsubsection*{Learning Halfspace Problem}

We consider a popular problem in learning theory : \emph{learning a halfspace in the presence of malicious errors}, as described in 
\cite{1997_ABSS}.

The input to the learner consists of a set of $k$ points in $\real^m$, each labeled with a $+$ (positive examples of a concept) or 
$-$ (negative examples of a concept). The learner's output is a hyperplane, $\braket{\vec{a},\vec{x}} = b$, where 
$\vec{a} \in \real^m$ and $b\in \real$. The hypothesis \emph{correctly classifies} a point $\vec{y}$ marked $+$ (or $-$) if it
satisfies $\braket{\vec{a},\vec{y}} > b$ (or $\braket{\vec{a},\vec{y}} < b$ respectively). Else, it \emph{misclassifies} the 
point.

Finding a hypothesis that minimizes the number of misclassifications is the \emph{open hemispheres problem}, which is $\NP$-hard.
The \emph{error} of the algorithm is the number of misclassifications by its hypothesis, and the \emph{noise} of the sample is the 
error of the best possible algorithm. The \emph{failure ratio} of the algorithm is the ratio of the error to noise.

The $\lhp$ can be formulated in the following way.

\begin{defn}[\textbf{Learning Halfspace Problem ($\lhp_c$)} \cite{1997_ABSS}]

Given a set of linear inequalities in $n$ variables over $\real$ and an integer $d$, distinguish between the
YES instance when there exists an assignment to the $n$ variables which does not satisfy at most $d$ inequalities, and the
NO instance when every assignment to the $n$ variables does not satisfy at least $c \cdot d$ inequalities.

\end{defn}

\section{Hardness result for $\ssat$}
\label{sec:lc2ssat}

We now prove the hardness of $\ssat$ (Theorem \ref{thm:ssat}) and $\ssati$ (Theorem \ref{thm:ssati}). 
In Section \ref{subsec:ssat1} we reduce Label Cover ($\lc$) to $\ssat$ and then 
appropriately adopt this reduction for $\ssati$ in Section \ref{subsec:ssati}.

Let $\mathcal{G}'=(G'=(A',B',E'),\Sigma_A,\Sigma_B,\Pi')$ be an $\lc$ instance obtained after applying the Projection Games 
Conjecture (Conjecture \ref{conj:pgc}). Assume it has size $N'$, soundness $s \geq 1/N'^{\beta}$ (for some $0<\beta<1$) and alphabet size 
$\poly(1/s)$.
We assume without loss of generality that the $\lc$ instance is bi-regular \cite{2010_MR,2012_M}. That is, every $A$ vertex has the 
same degree $D_L$ (which we call the \emph{left degree} or \emph{A-degree}) and every $B$ vertex has the same degree $D_R$ (which we
call the \emph{right degree} or \emph{B-degree}).

We use the following result from \cite{2012_M} which relates the soundness error to agreement soundness error. 
\begin{prop}[\cite{2012_M} (re-phrased)]
 Let $D_B \geq 2$ be a prime power, $D_R$ be a power of $D_B$ and $\epsilon >0$. From an $\lc$ instance with soundness error 
 $\epsilon^2D_B^2$ and $B$-degree $D_R$, we can construct an $\lc$ instance with agreement soundness error $2\epsilon D_B^2$ and 
 $B$-degree $D_B$. The transformation preserves the alphabets. The size is raised to a constant power.
\end{prop}

The running time of this reduction is polynomial in the size of the initial LC instance.
Thus after applying this lemma, we can assume we have an $\lc$ instance $\mathcal{G}=(G=(A,B,E),\Sigma_A,\Sigma_B,\Pi)$ with size
$N=N'^{\gamma}$ (for some constant $\gamma>1$), right degree $D_B$ (constant prime power), left degree $D_A$ and 
agreement soundness error $\agreesound=\frac{2s}{\epsilon}=2D_B\sqrt{s}$, where $\epsilon=\frac{\sqrt{s}}{D_B}$. Expressing in terms of $N$, we can write $\sound\geq 1/N^{\beta/\gamma}$. Here we note that $0<\beta<1$ and $\gamma>1$, thus $\sound\geq 1/N^c$ for some $0<c<1$.
Let for each edge $e\in E$, $\pi_e$ is a $p-to-1$ projection where $p \leq |\Sigma_A|$.

Our proof works even without the bi-regularity condition. What is crucial is the fact that degree of each $B$-vertex is bounded by some large enough constant. However, for convenience, we assume we have a bi-regular graph.

\subsection{Reduction from $\lc$ to $\ssat$}
\label{subsec:ssat1}

We reduce the above $\lc$ instance $\mathcal{G}$ to a $\ssat$ instance $\mathcal{I}=(V,\Psi,\mathcal{R}_{\Psi})$ as follows. 
To each $A$-vertex $a$ we associate a variable $\vect{a}$ i.e. $|V|=|A|$. To each $B$-vertex $b$ we associate a test $\psi_b$ i.e. $|\Psi|=|B|$. The variables in a test $\psi_b$ are the neighbors of $b$ in $A$. 
Thus each test has $D_B$ variables and each variable appears in $D_A$ tests. 

\textbf{Values of variables : } 
Without much loss of generality we assume that $\Sigma_A$ is in bijective correspondence with some field $\fld$, which is the range
of the variables in $V$. We use the letters $x$ and $y$ (with subscript and superscript as required) for the elements of $\Sigma_A$ 
(or $\fld$) and $\Sigma_B$  respectively. 

\textbf{Satisfying assignments for tests : }
Consider a $\psi_b\in \Psi$.
For each label $y \in \Sigma_B$ such that it has at least one pre-image in each of $b$'s neighbors in $A$, consider the following 
tuples:
\begin{eqnarray}
 \mathcal{R}_y(\psi_b) = \{(x_1,\ldots,x_{D_B}) : x_j \in \pi_{e}^{-1}(y) \text{ where } e=(a_j,b) \text{ and } a_j 
 \text{ is the } j^{th} \text{ neighbor of } b \} \nonumber
\end{eqnarray}

Thus the total set of satisfying assignments for $\psi_b$ is :
$\quad
 \mathcal{R}(\psi_b) = \bigcup_{y \in \Sigma_B} \mathcal{R}_y(\psi_b). \quad
$
And cardinality of this set is at most $|\Sigma_B|p^{D_B}$, which is polynomially bounded by $|V|$, since we assumed the Projection Games Conjecture is true. Each satisfying assignment $r$ is a tuple of size $D_B$ consisting  of values of some variables, as explained in Section \ref{sec:ssat}. So we sometimes use the word ``co-ordinate'' which implies the corresponding variable. From here on, when we say that a value in $r$ is assigned (or non-assigned) we imply that the value is assigned (or non-assigned) to its respective variable. 

\subsubsection*{Completeness}

\begin{lemma}

If there exists a labeling that satisfies all the edges in $G$ then there exists a consistent natural super-assignment 
satisfying all the tests in $\Psi$.

 \label{lem:complete}
\end{lemma}

\begin{proof}
 Suppose there is a labeling $\varphi_A: A \to \Sigma_A$, and $\varphi_B: B \to \Sigma_B$ that satisfies all edges. Then for any 
 $b \in B$, let 
 $ \quad 
 r_b := (\varphi_A(a_1), \ldots, \varphi_A(a_{D_B})), \quad
 $
 where $a_1, \ldots, a_{D_B}$ are the neighbors of $b$.
 Since the labeling ($\varphi_A$, $\varphi_B$) satisfies all edges, 
 we have that $r_b \in \mathcal{R}_{\varphi_B(b)}(\psi_b)$, and hence $r_b \in \mathcal{R}(\psi_b)$. 
 
 Consider the super-assignment for the resulting $\ssat$ instance that sets for all 
 $b \in B$, $\vec{S(\psi_b)}[r_b]=1$ and $\vec{S(\psi_b)}[r']=0$ for all $r'\neq r_b$. 
 This assignment is natural by definition and it is consistent since for each $\vect{a}\in V$ and for each neighbor $b$ of $a$, 
 we have that $\pi_{\vect{a}}(\vec{S(\psi_b)})[\varphi_A(a)] = 1$ and $\pi_{\vect{a}}(\vec{S(\psi_b)})[\alpha] = 0$ for any $\alpha \neq \varphi_A(a)$. 
 Thus we get a consistent natural super-assignment.

\end{proof}

 \subsubsection*{Soundness}
 
 We can make some observations about the structure of the $\ssat$ instance we constructed. These are not essential for our soundness
 proof, so we have moved these to \ref{app:ssat} and only state the following result. 
 \begin{lemma}[Corollary \ref{cor:app_assign1} in \ref{app:ssat}]
 
 For each test with non-zero norm, in the set of non-zero weighted assignments either there exists at least one assignment such that
 it has at least two variables with assigned values or all its assignments have exactly one variable with assigned value.
 
 \label{lem:assign1}
\end{lemma}
 Here we note that the non-triviality condition does not guarantee the existence of at least one variable with assigned value in ``each'' assignment.
 
 \begin{lemma}	

Let $D_B$ is a constant prime power such that $N$ is a power of $D_B$ and $0< c < 1$.
Let $s \geq 1/N^c$ and $\ell' < \frac{1}{\sqrt{2D_B} s^d}$, where $d<1/4$. Assume $\listsound = \sqrt{s}D_B\ell'^2$ \footnote{Here we note that we have an instance of LC with agreement soundness error $\agreesound=\frac{2s}{\epsilon}=2D_B\sqrt{s}$. By Lemma \ref{lemm:list_soundness}, for a list size $\tilde{\ell}$, it has list agreement soundnes $(\tilde{\ell},\listsound)$ where $\listsound=\agreesound \tilde{\ell}^2$. The bounds in the lemma have been chosen accordingly.}.

If $\mathcal{G}$ has list agreement soundness error $(\ell',\listsound)$ then every non-trivial consistent super-assignment for $\mathcal{I}$ has norm at least $g = N^{c'}(1-2\listsound)$, where $c'< d$.

 \label{lem:sound1}
\end{lemma}

\begin{proof}
We prove the contrapositive. Let $\mathcal{I}$ has a non-trivial consistent super-assignment $S$ of norm at most $g$. Then we prove that there exists a labeling such that for at least $\listsound$ fraction of $B$-vertices, the $A$-vertices do not totally disagree.

 If the average norm is at most $g$ then by Markov's inequality there exists at least $\listsound$ fraction of tests for which the norm, and hence the number of non-zero weight satisfying assignments, is at most $g_1=N^{c'}(1-\listsound)$.
 Let us denote this set of tests by $\Psi'(\subseteq \Psi)$. For any variable $\vect{a}$ we denote its set of assigned values by $\good_{\vect{a}}$ and the non-assigned values by $\ngood_{\vect{a}}$. For any assignment $r$ if $r|_{\vect{a}} \in\good_{\vect{a}}$ then we call the corresponding co-ordinate ``good''.
 Since $S$ is non-trivial each test must have at least one non-zero weight assignment with at least one good co-ordinate.

 We now define an assignment $\widehat{\phi_A}:A\rightarrow \binom{\Sigma_A}{\ell'}$ to the $A$-vertices. Let us call this the procedure of List-Construction.
 \begin{enumerate}
    \item For each variable we include all its assigned values in its list. \\
    \item For any test $\psi\in\Psi'$ if there are no assignment with at least two good co-ordinates, then we consider any one
    assignment with one good co-ordinate (existence of such an assignment is guaranteed by the non-triviality condition)
    and include each of the non-assigned values with probability $p=\min\{1,\frac{g_1}{D_A}\}$ in the lists of the 
 corresponding variables. \\
 (Lemma \ref{lem:assign1} exerts that in such a case each assignment will have exactly one good co-ordinate.)
 \end{enumerate}
 
 \paragraph{Expected list size}
 
 Consider a variable $\vect{a}$ (or corresponding vertex) and let $N_{\vect{a}}=|\good_{\vect{a}}| (\leq g_1)$ be the number of values assigned to it.
 By step (2) of List-Construction, from each of the $D_A$ test that $\vect{a}$ appears in, one non-assigned value can be included in the list with probability $p$.
 If $L_{\vect{a}}$ is the variable for the size of the list of labels for $a$, then
 \[
  \E[L_{\vect{a}}] \leq N_{\vect{a}} + D_A\cdot \frac{g_1}{D_A} \leq 2 g_1
 \] 
 This is true for every variable. Thus expected list size is at most $2g_1$ and with high probability it remains bounded by $\ell'$. By Markov's inequality, the probability of a variable (or corresponding vertex) having list size more than $\ell'$ is at most $\frac{2g_1}{\ell'}$. This probability is $\poly(\frac{1}{N})$ for small enough $c'$. Let $\mathcal{V}$ be the set of vertices with list size bounded by $\ell'$ and $\mathcal{V}'$ is the set of those vertices with list size more than $\ell'$. Each $\vect{v}\in\mathcal{V}'$ has $N_{\vect{v}}$ assigned values and rest non-assigned values. From each of its $D_A$ neigbours each such vertex can get at most one non-assigned value (by step 2). So the maximum number of non-assigned values is at most $D_A$. We select any $\ell'-N_{\vect{v}}>g_1$ of these values and discard the rest. This ensures that the list size for each vertex remains bounded by $\ell'$. Now  for each vertex $\vect{v}\in\mathcal{V}'$, probability ($p_v$) of including each non-assigned value is at least $\frac{g_1}{D_A}\cdot\frac{g_1}{D_A}>\frac{g_1^2}{D_A^2}$. 
 For the lists with size shorter than $\ell'$ we add dummy variables. This ensures that every list has size equal to $\ell'$.
 
 Now consider a test $\psi_1\in\Psi'$ with $\nzassign(\psi_1)=\{r_1,r_2,\ldots r_{g'}\}$ as the set of non-zero weighted assignments, where $g'\leq g_1$. Let $P_0(\psi_1)$ be the probability that for each $r\in\nzassign(\psi_1)$, not a single value appears in the list of its corresponding variable. By Lemma \ref{lem:assign1}, we can say $P_0(\psi_1)=0$ by Step 1 of List-Construction.
 (We can come to similar conclusion simply by noting the non-triviality condition, without using Lemma \ref{lem:assign1}.) Let $P_{\geq 2}(\psi_1)$ is the probability of the event that for any $r\in\nzassign(\psi_1)$ there exists at least two values that
 appear in the list of their variables. Here there can be two cases.
 
 In Case (i) there exists at least one $r\in\nzassign(\psi_1)$ such that it has at least two good co-ordinates. Then $P_{\geq 2}(\psi_1)=1$, by Step 1 of List-Construction.
 
 In Case (ii) for each $r\in\nzassign(\psi_1)$ there is at most one good co-ordinate. Then by Step 2 of List-Construction we know that there is an $r\in\nzassign(\psi_1)$ from which we took some non-assigned values. The probability that only one value is in the list of its corresponding variable is equal to the probability that none of the $D_B-1$ non-assigned values have been included and we only took the assigned value. This probability is given as follows.
 \begin{eqnarray}
  P_1(r)\leq\sum_{k=0}^{D_B-1}\binom{D_B-1}{k}\left(1-\frac{g_1}{D_A}\right)^{D_B-1-k}\left(\frac{2g_1}{\ell'}\left(1-\frac{g_1^2}{D_A^2}\right)\right)^k   \nonumber
 \end{eqnarray}
In the above equation we take all possible combinations where $k$ ($=0,\ldots,D_B-1$) variables are from $\mathcal{V}'$ and the rest from $\mathcal{V}$, as described before. Thus
\begin{eqnarray}
 P_1(r)\leq\left(1-\frac{g_1}{D_A}+\frac{2g_1}{\ell'}-\frac{2g_1^3}{D_A^2\ell'}\right)^{D_B-1}  \nonumber
\end{eqnarray}
 and $P_{\geq 2}(\psi_1)\geq 1-P_1(r) > \frac{c''g_1}{D_A}$, where $c''$ is a positive constant.
 
 Now consider test $\psi_2\in\Psi'$ with $\nzassign(\psi_2)$ being the set of non-zero weighted assignments. We calculate a bound on $P_{\geq 2}(\psi_2|\psi_1)$, which is the probability that $\exists r_2\in\nzassign(\psi_2)$ such that at least two of its values appear in the list of their corresponding variables, conditioned on the fact that such an event happens for $\psi_1$. Here we observe that there is a correlation between the tests in terms of the variables shared. But to calculate the required probability what is important is the fact that how the values have been selected in the list of the variables. This is determined by the List-Construction procedure. Due to consistency the assigned values of the shared variables are same and we include all of them in the list. Thus if $\exists r_2\in\nzassign(\psi_2)$ such that it has at least two good co-ordinates then $P_{\geq 2}(\psi_2|\psi_1)=1$.
 
 Now let us consider the case when for each $r_2\in\nzassign(\psi_2)$ there is at most one good co-ordinate. By Step 2 of List-Construction we know $\exists r_2\in\nzassign(\psi_2)$ from which we took one assigned value and few non-assigned values. Probability of including these non-assigned values in this step is not affected by other tests. We sort of ``look into'' each test separately and do the sampling independently. Even during discarding values from longer (than $\ell'$) lists we do not consider the tests from which these values have been taken. So the probability, $P_1(r_2|\psi_1)$, that only one value is in the list of its corresponding variable can be written as follows.
 \begin{eqnarray}
  P_1(r_2|\psi_1)\leq\left(1-\frac{g_1}{D_A}+\frac{2g_1}{\ell'}-\frac{2g_1^3}{D_A^2\ell'}\right)^{D_B-1}  \nonumber
\end{eqnarray}
 Thus $P_{\geq 2}(\psi_2|\psi_1)\geq 1-P_1(r_2|\psi_1) > \frac{c''g_1}{D_A}$, where $c''$ is a positive constant. Similar arguments and bound hold for $P_{\geq 2}(\psi_i|\psi_1,\ldots,\psi_{i-1})$.

 Let $P_{\geq 2}(\Psi')$ be the probability that for each test $\psi\in\Psi'$ there exists at least one assignment such that at least two of its values appear in the list of their corresponding variable. Then
 \[
P_{\geq 2}(\Psi')=\prod_{\psi_i\in\Psi'} P_{\geq 2}(\psi_i|\psi_1,\ldots,\psi_{i-1})\geq
 \Big(\frac{c''g_1}{D_A} \Big)^{\listsound |B|} > 0.
 \]
 
 Thus there exists a labeling of the $A$-vertices (variables) such that for every $b\in\Psi'$ (tests), we have at least two 
 vertices $a_1,a_2$ in $A$ such that $\sigma_1 \in \widehat{\phi_A}(a_1),\quad \sigma_2 \in \widehat{\phi_A}(a_2)$ and 
 $\pi_{e_1}(\sigma_1)=\pi_{e_2}(\sigma_2)$, where $e_1=(a_1,b), e_2=(a_2,b) \in E$, i.e. both the vertices agree (values appear in
 the same assignment). Hence we do not have an agreement soundness of $(\ell,\listsound)$.
 
 \end{proof}

 Thus we prove Theorem \ref{thm:ssat}: $\quad$
 \emph{Assuming the Projection Games Conjecture $\ssat$ is $\NP$-hard for $g=n^c$ for some constant $c>0$.}

 \begin{remark}
  By Lemma \ref{lem:sound1} the constant $c$ in the above theorem is less than $1/2$. This is because the given $\lc$ instance has $\listsound \propto \agreesound\ell^2$ and $\agreesound\propto\sqrt{s}$. Since $\listsound <1$, so $g < \sqrt{N}$, where $N$ is the size of the $\lc$ instance.
 \end{remark}

 \subsection{Reduction from $\lc$ to $\ssati$}
 \label{subsec:ssati}

 Given an $\lc$ instance $\mathcal{G}$ we construct an $\ssati$ instance $\mathcal{I}$, as done in Section \ref{subsec:ssat1}. 
 The completeness proof is also similar to Lemma \ref{lem:sound1}, so we do not repeat it again here.
 
 \subsubsection*{Soundness}
 
 We need to show that a ``NO'' instance of $\lc$, i.e. when it has a list-agreement soundness of $(\ell',\listsound)$, maps to a ``NO'' instance of $\ssati$, i.e. when every consistent not-all-zero super-assignment has norm
 at least $g$. When the norm is between $1$ and $g$ then any outcome (YES/NO) is fine.
 
 If we assume that the constructed $\ssati$ instance has a consistent non-trivial solution of norm at most $g$, then similar to 
 Lemma \ref{lem:sound1} we can prove that there exists a labeling such that the given $\lc$ instance does not have a list agreement 
 soundness error $(\ell',\listsound)$.
 The only difference is that since here $\ell_{\infty}$ norm is at most $g$, so norm of every test remains bounded by $g$.
 Note that a non-trivial solution is also not-all-zero.
  \begin{lemma}	

Let $D_B$ is a constant prime power such that $N$ is a power of $D_B$ and $0< c < 1$.
Let $s \geq 1/N^c$ and $\ell' < \frac{1}{\sqrt{2D_B} s^d}$, where $d<1/4$. Assume $\listsound = \sqrt{s}D_B\ell'^2$.

If $\mathcal{I}$ has a consistent non-trivial super-assignment of $\ell_{\infty}$ norm at most $g=N^{c'}$ then $\mathcal{G}$ has 
list agreement soundness error $(\ell',\listsound)$, where $c'< d$.

 \label{lem:soundi1}
\end{lemma}
Now suppose the constructed $\ssati$ instance of norm at most $g$ does not have a non-trivial solution, i.e. every variable does not
have at least one value assigned, but the solution is consistent and not-all-zero. Let $\Psi'\subseteq\Psi$ is the set of tests with non-zero norm and $V'\subset V$ is the set of variables with no assigned values, i.e. for every $\vect{a}\in V'$ we have $\good_{\vect{a}}=\emptyset$. When a value has projection weight $0$, it can be due to the weights of the super-assignments (and how they cancelled)
or because the value never appeared in any super-assignment (e.g. if the projection functions are partial functions).
Thus in the $\ell_{\infty}$ case even if consistency is maintained, for variables with no assigned values, we cannot say that one 
particular value appears in all its tests.
Then we can change the List-Construction procedure of Lemma \ref{lem:sound1} as follows. Lemma \ref{lem:assign1} remains true even in the $\ell_{\infty}$ norm.
\begin{enumerate}
 \item For each variable we include all its assigned values in its list.
 
 \item For any test in $\psi\in\Psi'$ if each of its non-zero assignment has exactly one good co-ordinate, then we consider any one 
 such assignment and include each of the non-assigned values with probability $p=\min\{1,\frac{g}{D_A}\}$ in the list of the 
 corresponding variables. Once values have been included we consider the tests as ``marked''.
 
 \item If $\Psi\setminus\Psi'\neq\emptyset$ then we do the following sequentially. Consider any variable in $V'$ (say $\vect{a_1}$) 
 and let $\Psi'_{\vect{a_1}}\subseteq\Psi\setminus\Psi'$ be the tests in which it appear. Start with any test in $\Psi'_{\vect{a_1}}$ and take any one of its assignment. We include the value of $\vect{a_1}$ in its list and the values of the other variables (in the lists of the respective variables) with probability $p$. We consider this test ``marked'' and go
 to the next one in $\Psi'_{\vect{a_1}}$. If there are any assignment such that the value of $\vect{a_1}$ has already been included in the list, then we consider it. Else we take any assignment and include the value of $\vect{a_1}$ in its list. The values of the other variables are included in their lists with probability $p$. This continues till we have taken at most $g$ values of $\vect{a_1}$ (we call the values obtained from this step as
 ``marked values''). Let $\Psi''\subset\Psi\setminus\Psi'$ be the set of tests marked in this way. We repeat this procedure again
 with another variable in $V'$ and the remaining ``unmarked'' tests. This continues till we exhaust all variables in $V'$.
\end{enumerate}

The expected list size for variables with assigned values remain at most $2g$, similar to Lemma \ref{lem:sound1}. For each
variable $\vect{a_1}\in V'$ there can be at most $g$ ``marked'' values. In each of the $D_A$ tests that $\vect{a_1}$ appears one ``unmarked'' value (apart from the marked one) can be in the list with probability $p$. 
So expected list size is at most $g+D_A\cdot \frac{g}{D_A} = 2g$. Similar to Lemma \ref{lem:sound1} here also we separate the variables into two sets $\mathcal{V}$, where the list size is bounded by $\ell'$ and $\mathcal{V}'$ where the list size is more than $\ell'$. We can discard some non-assigned or unmarked values and ensure that the list size remains equal to $\ell'$ for each variable.

In Step (3) of List-Construction we see that each variable $\vect{a_1}\in V'$ can mark at least $g$ tests in $\Psi'$. Each variable
with assigned values can mark $D_A$ tests each. By the condition of not-all-zero there is at least one variable with assigned values.
Thus at least $g|A|$ tests can be marked. So $\listsound=\frac{g|A|}{|B|}=\frac{gD_B}{D_A}$.
Next, we can argue in the same way as in Lemma \ref{lem:sound1} that for at least $\listsound$ fraction of the $B$-vertices there
will not be total disagreement.

Now, one crucial thing here is, for the parameters to make sense, $gD_A \approx \frac{1}{\sqrt{s}}$. Or, the projection functions
are such that more tests can be marked in Step (3). For example if they are total functions. Or they are partial functions but there exists a set of variables $V'' \subseteq V$ such that we can find a set of values $\mathcal{V}_{\vect{a}}$ ($\vect{a}\in V''$) of cardinality at most $g$ such that for at least $\listsound$ fraction of tests there exists at least one pre-image in that set i.e. for each such test there exists some assignment $r$ such that $r|_{\vect{a}}\in\mathcal{V}_{\vect{a}}$.

Thus we can state the following lemma.
\begin{lemma}
 Let $D_B$ is a constant prime power such that $N$ is a power of $D_B$ and $0< c < 1$.
Let $s \geq 1/N^c$ and $\ell' < \frac{1}{\sqrt{2D_B} s^d}$, where $d<1/4$. Assume $\listsound = \sqrt{s}D_B\ell'^2$.

If $\mathcal{I}$ has a consistent not-all-zero (but not non-trivial) super-assignment of $\ell_{\infty}$ norm at most $g=N^{c'}$ ($c'< d$), then
$\mathcal{G}$ has list agreement soundness error $(\ell',\listsound)$ if either \\
(i) $gD_A \approx \frac{1}{\sqrt{s}}$; OR $\quad$
(ii) For most variables we can find a set of values of cardinality at most $g$ such that for at least $\listsound$ fraction of tests
(with that variable) there exists at least one assignment with values in that set.

\label{lem:soundi2}
\end{lemma}
This proves Theorem \ref{thm:ssati}: $\quad$
\emph{Assuming the Projection Games Conjecture $\ssati$ is $\NP$-hard for $g=n^c$ for some constant $c>0$.} Again from Lemma \ref{lem:soundi2} and the preceding discussion we get $c<1/2$.


\section{Applications : Reduction from SSAT to other problems}
\label{sec:app}

\subsection{Complexity of lattice problems}

Dinur et.al. \cite{2003_DKRS} reduced $g-\ssat$ to $g-\sis$ and $g-\cvp$. 
In a separate paper Dinur \cite{2002_D} reduced $g-\ssati$ to $g-\svp_{\infty}$.
Thus using Theorem \ref{thm:ssat} and \ref{thm:ssati} we have the following corollaries.

\begin{cor}
Assuming the Projection Games Conjecture $\sis$ is $\NP$-hard to approximate within a factor $g=N^c$, for some constant $c >0$.
\label{cor:sis_pgc}
\end{cor}

\begin{cor}
Assuming the Projection Games Conjecture $\cvp$ on $N$-dimensional lattice is $\NP$-hard to approximate within a factor $g=N^c$, 
for some constant $c >0$.
\label{cor:cvp_pgc}
\end{cor}

\begin{cor}
Assuming the Projection Games Conjecture $\svp_{\infty}$ on $N$-dimensional lattice is $\NP$-hard to approximate within a factor 
$g=N^c$, for some constant $c >0$.
\label{cor:svp_pgc}
\end{cor}

From the discussion in Section \ref{sec:lc2ssat} and keeping in mind the blow-up in size we get from the reduction of $\ssat$ to $\sis$ (\ref{app:ssat2sis}) we can conclude that the constant $c$ in the exponent in each of the above corollaries is less than $1/2$.
Thus these results do not violate the polynomial hierarchy \cite{2000_GG}.
\subsection{Complexity of Nearest Codeword Problem ($\ncp$)}

Here we give an approximation factor preserving reduction from $\ssat$ to $\ncp$ and prove the following using the $\ssat$ Theorem \cite{2003_DKRS}.
\begin{theorem}
Approximating $\ncp$ within factor $n^{c/\log \log n}$ is NP-hard.
\label{thm:ncp}
\end{theorem}

\begin{proof}
Let $\mathcal{I}=\langle \Psi=\{\psi_1,\ldots,\psi_{n}\}, V=\{\vect{v}_1,\ldots,\vect{v}_m\},\{\mathcal{R}_{\psi_1},
\ldots,\mathcal{R}_{\psi_{n}}\} \rangle$ be an $\ssat$ instance. From this we efficiently construct a matrix $\vec{B'}$  ($\sis$ 
matrix) and a target vector $\vec{t'}$, as given in \cite{2003_DKRS}. Here we give a brief description of this matrix and the 
relevant facts that we require. 
For completeness we give the reduction from $g-\ssat$ to $g-\sis$ and prove these facts in \ref{app:ssat2sis}.

$\vec{B'}$ is a $0-1$ matrix and its dimension is $n' \times m'$ where $n',m'\in\poly(n)$. 
$\vec{t}'$ is an all-$1$ column vector of length $n'$.
If the $\ssat$ instance has a natural consistent super-assignment (YES) then there exists a coefficient vector $\vec{z}$ such that
$\vec{B'z}=\vec{t'}$ and $\|\vec{z}\|_1=n$. If every non-trivial consistent super-assignment of the $\ssat$ instance has norm 
greater than $g$ (NO) then for all coefficient vectors $\vec{z}$ with $\vec{B'z}=\vec{t'}$ we have $\|\vec{z}\|_1> gn$.								
Let $D > gn$ be some integer. We construct the matrix $\vec{A} \in \{0,1\}^{n'D+m' \times m'}$ for $\ncp$ as follows. 

We can think of $\vec{A}$ as consisting of two parts -  the first $n'D$ rows in the upper part ($\vec{A'}$, say) and the rest $m'$ 
rows in the lower part. In the upper part $\vec{A}[ik,j]=\vec{B}'[i,j]$ for $1\leq i \leq n', 1\leq k \leq D, 1\leq j \leq m'$. 
The lower part is the identity matrix $\vec{\id}^{m' \times m'}$.

The target vector $\vec{t}$ is as follows : $t_{ik} = 1$ for $1\leq i \leq n', 1\leq k \leq D$. Let us call these rows as 
$\vec{t}''$. The rest of $\vec{t}$ is $0$. 

Let $q$ be a prime number greater than $g$ times $\max \{n',m'\}$. 
The $\ncp$ instance obtained by reduction is $(\vec{A},\vec{t},n)$ over $\mathbb{F}_q$.  
For any coefficient vector $\vec{z}$ we have $\|\vec{Az}-\vec{t}\|=\|\vec{A'z}-\vec{t}''\|+\|\vec{z}\|$.

\textbf{Completeness :} There exists $\vec{z}$ such that $\vec{B'z}=\vec{t}'$ and $\|\vec{z}\|_1=n$. 
This implies $\|\vec{Az}-\vec{t}\| = \|\vec{z}\|_1=n$ and $\ncp$ oracle outputs YES.

\textbf{Soundness :} For all coefficient vectors $\vec{z}$ where $\vec{B'z}=\vec{t}$ we have $\|\vec{z}\|_1> gn$.
In the $\ncp$ instance if $\vec{A'z}-\vec{t}'' \neq \vec{0}$ then it implies that at least $D$ co-ordinates are non-zero, 
else it implies $\vec{B'z}=\vec{t}'$
In both the cases $\|\vec{Az}-\vec{t}\| > gn$ and the $\ncp$ oracle outputs NO.	
\end{proof}

We can get an improved hardness of approximation factor using Theorem \ref{thm:ssat}.
\begin{cor}
	Assuming the Projection Games Conjecture, approximation of $\ncp$ up to a factor of $n^c$ is $\NP$-hard, for some constant $c>0$.
	\label{cor:ncp_pgc}
\end{cor}

\subsection{Complexity of Learning Halfspaces Problem ($\lhp$)}
\begin{theorem}
Approximating the minimum failure ratio of Learning Halfspace Problem ($\lhp$) within a factor of $n^{c/\log\log n}$ is $\NP$-hard.
\label{thm:lhp}
\end{theorem}

\begin{proof}
From an $\ssat$ instance 
$\mathcal{I}=\langle \Psi=\{\psi_1,\ldots,\psi_{n}\}, V=\{\vect{v}_1,\ldots,\vect{v}_m\},\{\mathcal{R}_{\psi_1},
\ldots,\mathcal{R}_{\psi_{n}}\} \rangle$, we construct an $\sis$ instance $(\vec{B'},\vec{t'},d)$ as in \cite{2003_DKRS} 
 and derive inequalities, somewhat simiar to \cite{1997_ABSS}. The theorem then follows from the $\ssat$ Theorem \cite{2003_DKRS}.
 
 A brief description of the $\sis$-matrix $\vec{B'}$ and the relevant
 facts for this proof have been given in Theorem \ref{thm:ncp}. More details can be found in \ref{app:ssat2sis}.
Observe that an $\sis$ instance $(\vec{B}',\vec{t'},d)$ can be viewed as a set of linear equations 
$\vec{B'z}-\vec{t'}=\vec{0}$ for some variable vector $\vec{z}$.  
Note that while learning a halfplane $\braket{\vec{a},\vec{x}}=b$, the unknown variables are $\vec{a}$ and $b$.
So it is a homogeneous system and the coefficient of one of the variables (here $b$) is always $\pm 1$. We use a standard technique 
for homogenization of the $\sis$ equations- use a new variable $y$ and replace every constant $c$ by $cy$.

Let $U=n^{c/\log\log n}|\Psi|$ and $d=|\Psi|$. Following are the linear inequalities for $\lhp$ instance.
\begin{enumerate}
 \item We make $U$ copies of the linear inequality :    $ \quad \dfrac{-y}{U}<\delta< \dfrac{y}{U} $.
			
  \item For each equation from $\sis$ of the form $\sum_{i=1}^na_ix_i=c$, we make $U$ copies of each of the following two 
  inequalities:
      \[  \sum_{i=1}^na_ix_i-cy+\delta > 0 , \qquad  \sum_{i=1}^na_ix_i-cy-\delta < 0\]
		
  \item For every variable make $U$ copies of each of the following :
			\[x_i-2y<0,\hspace{3mm} x_i+2y>0 \]
			
  \item For every variable make a copy of the following :
			\[x_i+\delta > 0 , \qquad
			x_i-\delta < 0\]		
  \item Make $U$ copies of the inequality : $ \quad y > 0$
\end{enumerate} 

\textbf{Completeness : } There exists a $\{0,1\}$ coefficient vector (variables $\vec{x}$) with norm $d$ that satisfies the linear
equations. In the $\lhp$ instance assign the same values to the $x_i$ variables and put $y=1$ and $\delta$ a very small number
tending to $0$. This will satisfy all the inequalities of type $1$, $2$ and $3$. Among inequalities of type $4$ only when $x_i=1$
the second inequality will not be satisfied. Thus the number of unsatisfied linear inequalities are $d$.
		 
\textbf{Soundness : } We give a contrapositive argument. 
Suppose there exists an assignment such that less than $U$ inequalities are unsatisfied. 
So the assignment must satisfy all inequalities of type $1,2$ and $3$. Now in the $\sis$ instance simply assign the variables the 
corresponding values divided by value of $y$. This implies the variables in $\sis$ get values from $\{-1,0,1\}$ and satisfies all 
the $\sis$ equations. Hence we have a solution with norm less than $U$ for $\sis$. This would imply the existence of a non-trivial
consistent super-assignment of norm at most $n^{c/\log\log n}$ for the $\ssat$ instance.
	 
\end{proof}

Using Theorem \ref{thm:ssat} we get the following corollary.
\begin{cor}
Assuming the Projection Games Conjecture, approximating the minimum failure ratio of $\lhp$ up to a factor of $n^c$ is $\NP$-hard, for
some constant $c>0$.
	\label{cor:lhp_pgc}
\end{cor}

From the discussion in Section \ref{sec:lc2ssat} the constant $c$ in the exponent in Corollary \ref{cor:ncp_pgc} and \ref{cor:lhp_pgc} is less than $1/2$.

\section*{Acknowledgement}
The author would like to thank Divesh Aggarwal and Rajendra Kumar for helpful discussions. The author would like to thank Dana Moshkovitz for clarifying some concepts about Projection Games Conjecture, via personal correspondence with Divesh Aggarwal. The author would like to thank anonymous reviewers for helpful comments. The author was partly funded by National University of Singapore during this research. Research at IQC was supported in part by the Government of Canada through Innovation, Science and Economic Development Canada, Public Works and Government Services Canada and Canada First Research Excellence Fund.


\appendix
\newtheorem{atheorem}{Theorem}
\newtheorem{alemma}{Lemma}
\newtheorem{aclaim}{Claim}
\newtheorem{acorollary}{Corollary}

\section{Some results about the $\ssat$ instance in Section \ref{subsec:ssat1}}
\label{app:ssat}

Here we prove some results about the $\ssat$ instance obtained in Section \ref{subsec:ssat1} by a reduction from an $\lc$ instance.
We give a quick recap of some relevant facts.

We are given an $\lc$ instance $\mathcal{G}=(G=(A,B,E),\Sigma_A,\Sigma_B,\Pi)$ with size $N$, right degree $D_B$ (constant prime 
power), left degree $D_A$. For each edge $e\in E$, $\pi_e$ is a $p-to-1$ projection where $p \leq |\Sigma_A|$.

We reduce $\mathcal{G}$ to a $\ssat$ instance $\mathcal{I}=(V,\Psi,\mathcal{R}_{\Psi})$ as follows. 
To each $A$-vertex $a$, we associate a variable $\vect{a}$, i.e. $|V|=|A|$. To each $B$-vertex $b$ we associate a test $\psi_b$, 
i.e. $|\Psi|=|B|$.The variables in a test $\psi_b$ are the neighbors of $b$ in $A$. 

\textbf{Values of variables : } 
Without much loss of generality we assume that the variables take values from a field $\fld$, which is in bijective correspondence 
to $\Sigma_A$. We use the letters $x$ and $y$ (with subscript and superscript as required) for the elements of $\Sigma_A$ (or $\fld$)
and $\Sigma_B$  respectively. 

\textbf{Satisfying assignments for tests : }
Consider a $\psi_b\in \Psi$.
For each label $y \in \Sigma_B$ such that it has at least one pre-image in each of $b$'s neighbors in $A$, consider the following 
tuples:
\begin{eqnarray}
 \mathcal{R}_y(\psi_b) = \{(x_1,\ldots,x_{D_B}) : x_j \in \pi_{e}^{-1}(y) \text{ where } e=(a_j,b) \text{ and } a_j 
 \text{ is the } j^{th} \text{ neighbor of } b \} \nonumber
\end{eqnarray}

Thus the total set of satisfying assignments for $\psi_b$ is :
$\quad
 \mathcal{R}(\psi_b) = \bigcup_{y \in \Sigma_B} \mathcal{R}_y(\psi_b) 
$.
And cardinality of this set is at most $|\Sigma_B|p^{D_B}$, which is polynomially bounded by $|V|$, assuming the Projection Games Conjecture is true.

Consider a test $\psi\in\Psi$ consisting of the variables 
 $\vect{a_1},\vect{a_2},\ldots,\vect{a_{D_B}}$.
 We can partition its set of satisfying assignments $\mathcal{R}(\psi)$ as follows: 
 
 For each $y\in\Sigma_B$, $\mathcal{R}_y(\psi)$ can be viewed as a $D_B$-dimensional array $\vec{M^y}$ where the $i^{th}$ dimension or 
 co-ordinate corresponds to variable $\vect{a_i}$. The number of values of $\vect{a_i}$ in its dimension is the number of pre-images 
 of $y$. Thus there can be at most $p$ values in each dimension. 
 With a slight abuse of notation, we denote array elements by $M^y_{x_1,\ldots,x_{D_B}}$, where $x_j\in\pi_e^{-1}(y)$, 
 $e=(a_j,b)$ and $a_j$ is the $j^{th}$ neighbor of $b$. 
 This is the weight of the corresponding assignment $(x_1,\ldots,x_{D_B})$.
 Thus the projection  of a super-assignment $\vec{S(\psi)}$ on a variable $\vect{a_i}$ for any pre-image of $y$ is :
 \[
  \forall x\in\pi_e^{-1}(y): \qquad \pi_{\vect{a_i}}\vec{(S(\psi))}[x] = \sum_{x_1,\ldots,x_{i-1},x_{i+1},\ldots,x_{D_B}} 
    M^y_{x_1,\ldots,x_{i-1},x,x_{i+1},\ldots,x_{D_B}}
 \]
 where the summation is over all the pre-images of $y$ for each variable $\vect{a_1},\ldots,\vect{a_{i-1}},\vect{a_{i+1}},\ldots 
 \vect{a_{D_B}}$.
 Note, since each edge constraint are functions, so the values for each variable that appear in one array cannot come in the other.
 In this case the arrays can be considered as ``disjoint'' or ``non-interfering''. While taking projection on a variable for any 
 value, it is sufficient to consider only the array in which this value appears, and not the other arrays.
 
 We can build these arrays in polynomial time and for convenience we can think of giving these arrays as input to the $\ssat$
 oracle, so that it can fill the arrays with weights according to some well-defined criteria.
 A co-ordinate of an array is \emph{good} (with respect to it) if at least one of the values is assigned to its
 corresponding variable, else it is \emph{bad}.
 
 We define the \emph{norm} of each array $\vec{M^y}$ as 
 \[
  \|\vec{M^y}\| = \sum_{x_1,\ldots,x_{D_B}} |M_{x_1,\ldots,x_{D_B}}|
 \]
 The norm of a super-assignment $\vec{S(\psi)}$ is:
 \[
  \|\vec{S(\psi)}\| = \sum_{y\in\Sigma_B} \|\vec{M^y}\|
 \]
(If the reduction is to $\ssati$ then $\|\vec{S(\psi)}\|_{\infty} = \max_{y\in\Sigma_B} \|\vec{M^y}\|$. )
 
 \begin{aclaim}
  With a consistent super-assignment if a test has an array with non-zero norm but all bad co-ordinates, then we can always find a
  consistent super-assignment with a lesser norm.
  \label{claim:0}
 \end{aclaim}

 \begin{proof}
 This follows from the ``disjoint''-ness of the arrays, as has been explained above. Since the weights cannot cancel from entries
 in other arrays, we can put all-zero weight in this array (with all bad co-ordinates). We get a 
 super-assignment of smaller norm, but consistency is maintained. 
 \end{proof}
 
 \begin{aclaim}
  In any array if any variable has all non-assigned values, i.e. any co-ordinate is bad, then the total sum of all the entries is 
  zero.
  \label{claim:1}
 \end{aclaim}
 
 \begin{proof}
  Let in array $\vec{M^y}$ the first variable $\vect{a_1}$ has all non-assigned values. Thus for each value $x$ of $\vect{a_1}$
  \[
   \pi_{\vect{a_1}}[x] = \sum_{x_2,\ldots,x_{D_B}} M^y_{x,x_2,\ldots,x_{D_B}} = 0.
  \]
  The sum of all array entries is
  \[
    \sum_{x_1,\ldots,x_{D_B}} M^y_{x_1,\ldots,x_{D_B}} = \sum_{x_1}\sum_{x_2,\ldots,x_{D_B}} M^y_{x_1,x_2,\ldots,x_{D_B}} = 0.
  \]
 \end{proof}

 \begin{aclaim}
 In an array with not-all bad co-ordinates, there can be either at least two good co-ordinates or one good co-ordinate with at 
 least two values assigned to the corresponding variable.
 
 If there is only one good co-ordinate then a consistent super-assignment can be made with a constant (one should
 suffice) number of non-assigned values in the other co-ordinates.
 
  \label{claim:2}
 \end{aclaim}

 \begin{proof}
  If possible let there be only one good co-ordinate corresponding to variable $\vect{a_1}$ (say) and it has only one assigned value,
  say $x$.
  Thus $\sum_{x_2,\ldots,x_{D_B}} M_{x,x_2,\ldots,x_{D_B}} = w \neq 0$ and 
  $\sum_{x_2,\ldots,x_{D_B}} M_{x',x_2,\ldots,x_{D_B}} = 0$, for all $x'\neq x$ (which appear in this matrix).
  
  Hence $\sum_{x_1,\ldots,x_{D_B}} M_{x_1,\ldots,x_{D_B}} = w \neq 0$, which cannot be true by Claim \ref{claim:1}.
  
  For the second part of the claim, without loss of generality let $\vect{a_1}$ be the co-ordinate with assigned values $1,2,\ldots, x$ and 
  corresponding projection weights $w_1,w_2,\ldots w_x$. The following conditions must hold:
  \[
   \sum_{x_2,\ldots x_{D_B}} M_{i,x_2,\ldots,x_{D_B}} = w_i, \qquad i=1,\ldots,x
  \]
  By Claim \ref{claim:1} $\sum_i w_i = 0$.
  
  Thus we can have a super-assignment as follows: Fix $\vect{a_2},\ldots, \vect{a_{D_B}}$ to some value $x_2,\ldots,x_{D_B}$ 
  respectively. Assign $M_{ix_2,\ldots,x_{D_B}} = w_i$, for all $i=1,\ldots, x$. It is easy to see that consistency is maintained 
  and this gives the minimum norm.
 \end{proof}

 \begin{aclaim}
If an array has more than one good co-ordinates then either (i) there exists at least one assignment with at least 
two assigned values or (ii) each assignment has one assigned value and there are only a constant number of non-assigned values 
for each variable.

 \label{claim:3}
\end{aclaim}

\begin{proof}
 Easy to see from Claim \ref{claim:2}.
\end{proof}

As a corollary of the above claims we can conclude that
\begin{acorollary}
 For each test with non-zero norm, in the set of non-zero weighted assignments either there exists at least one assignment such that it has at least
 two variables with assigned values or all its assignments have exactly one variable with assigned value.
 
 \label{cor:app_assign1}
\end{acorollary}

\section{$g-\ssat$ to $g-\sis$ reduction }
\label{app:ssat2sis}

In this section we give a brief outline of the approximation factor preserving reduction from $\ssat$ to $\sis$ given by Dinur et al.
\cite{2003_DKRS}. Given a $g-\ssat$ instance $\mathcal{I}=\langle\Psi=\{\psi_1,\ldots,\psi_{n}\},V=\{\vect{v}_1,\ldots,\vect{v}_m\},
\{\mathcal{R}_{\psi_1},\ldots,\mathcal{R}_{\psi_{n}}\} \rangle$ we construct a $g-\sis$ instance $\mathcal{S}=(\vec{B},\vec{t},d)$ as follows.

The \textbf{target vector} $\vec{t}$ is an all-$1$ vector and $d=|\Psi|=n$.

\textbf{The $\mathbf{\sis}$-matrix} $\vec{B}$ has a column for every pair $(\psi,r)$ where $\psi\in\Psi$ is a test and 
$r\in\mathcal{R}_{\psi}$ is a satisfying assignment for it. Thus there are $\sum_{i=1}^n |\mathcal{R}_{\psi_i}|$ columns.
It can be divided into two parts : the upper part consists of
\emph{consistency rows} to take care of consistency and the lower part consists of \emph{non-triviality rows} to take care of
non-triviality.

\textbf{Non-triviality rows : } There is a row for each test. In the row for $\psi$ all the columns associated with $\psi$ have $1$,
and all the other columns have $0$. Thus there are $|\Psi|$ non-triviality rows.

\textbf{Consistency rows : } There are $|\fld|$ rows for each pair of tests $\psi_i$ and $\psi_j$ and common variable $x$ shared by
them. If $a_{ij}$ is the number of variables shared by $\psi_i$ and $\psi_j$ then the number of consistency rows is
$\sum_{i,j} a_{ij}\cdot|\fld|$.
These rows contain a \emph{consistency-ensuring gadget} and only the columns for $\psi_i$ and $\psi_j$ will have non-zero 
values in these rows. 

The \textbf{consistency-ensuring gadget} for pair of tests $\psi_i$ and $\psi_j$ with common variable $x$ ensures that the 
super-assignments to these tests are consistent on $x$. It consists of a pair of matrices 
$\vec{G_1}^{|\fld|\times|\mathcal{R}_{\psi_i}|}$ and $\vec{G_2}^{|\fld|\times|\mathcal{R}_{\psi_j}|}$. The $|\fld|$ rows of each 
matrix correspond to the possible assignments for the variable $x$. The $r^{th}$ column in $G_1$ is the \emph{characteristic function}
of $r|_x$, i.e. it has $1$ on the value of $x$ in $r$, and $0$ everywhere else. For $G_2$ the $r'^{th}$ column is a ``sort of''
negation of the characteristic function of $r'|_x$, i.e. it has $0$ on the value of $x$ in $r'$ and $1$ everywhere else.

\paragraph{Correctness}
We show that the YES instance of $g-\ssat$ maps to the YES instance of $g-\sis$.
\begin{alemma}
 If there is a consistent natural super-assignment to the $g-\ssat$ instance $\mathcal{I}$ then there exists a solution of $\ell_1$
 norm $|\Psi|$ to the $g-\sis$ instance $\mathcal{S}$.
 \label{lem:ssat2sis_complete}
\end{alemma}

\begin{proof}
 Let $S$ be a consistent natural super-assignment. We will construct a solution $\vec{z}$ to the $g-\sis$ as follows: Note each 
 $\vec{S(\psi_i)}$ is a $|\mathcal{R}_{\psi_i}|$-length vector. $\vec{z}$ is a $\sum_{i=1}^n|\mathcal{R}_{\psi_i}|$-long vector
 consisting of the concatenation of the vectors $\vec{S(\psi_1)},\vec{S(\psi_2)},\ldots,\vec{S(\psi_n)}$.
 
 Since $S$ is natural, it assigns a $+1$ to exactly one assignment of every test. Thus the target vector is reached in the 
 non-triviality rows.
 
 To show that the target vector is reached in the consistency rows, consider the $|\fld|$ rows belonging to a pair of tests $\psi_i$
 and $\psi_j$ with common variable $x$. Let $\vec{S(\psi_i)}[r_1]$ and $\vec{S(\psi_j)}[r_2]$ be the single $1$'s in $\vec{S(\psi_i)}$
 and $\vec{S(\psi_j)}$ respectively. Since $S$ is consistent so $r_1|_x=r_2|_x$. By the construction of the gadget matrices in $\vec{B}$
 we see that the sum of these two columns gives an all-$1$ vector. So the target vector is reached in the consistency rows as well.
 
 $\|\vec{z}\|_1=\sum_{i=1}^n \|\vec{S(\psi_i)}\|_1 = |\Psi|$, since $\|S\|=1$.

\end{proof}

\paragraph{Soundness}
We need to show that a NO instance of $g-\ssat$ maps to a NO instance of $g-\sis$. Instead we give a contrapositive argument and
prove the following.
\begin{alemma}
If there exists a solution $\vec{z}$ of the $g-\sis$ instance $\mathcal{S}$ such that $\|\vec{z}\|_1\leq g|\Psi|$, then there exists 
a non-trivial consistent super-assignment $S$ of norm at most $g$ for the $g-\ssat$ instance $\mathcal{I}$.

 \label{lem:ssat2sis_sound}
\end{alemma}

\begin{proof}
 Given $\vec{z}$ we construct a super-assignment $S$ as follows : $\vec{z}$ is of length $\sum_{i=1}^n|\mathcal{R}_{\psi_i}|$.
 We break $\vec{z}$ into $|\Psi|$ pieces of length $|\mathcal{R}_{\psi_1}|,\ldots,|\mathcal{R}_{\psi_n}|$, one for each test
 $\psi\in\Psi$. We obtain a super-assignment of norm $\|S\|=\frac{1}{|\Psi|}\|\vec{z}\|_1 \leq g$.
 
 Since for each $\psi\in\Psi$, the target vector is reached in the $\psi^{th}$ row of the non-triviality rows, so
 \begin{eqnarray}
  \sum_{r\in\mathcal{R}_{\psi}} \vec{S(\psi)}[r]=1
  \label{eqn:sis1}
 \end{eqnarray}
and $S$ is non-trivial.

Let $\psi_i,\psi_j\in\Psi$ with common variable $x$. Consider the $|\fld|$ rows that correspond to $\psi_i,\psi_j,x$. In each of 
these rows the sum of the vectors is $1$, i.e. for any $f\in\fld$,
\begin{eqnarray}
 \sum_{r:r|_x=f} \vec{S(\psi_i)}[r] +\sum_{r:r|_x\neq f}\vec{S(\psi_j)}[r]=1
 \label{eqn:sis2}
\end{eqnarray}
Subtracting Eq.(\ref{eqn:sis1}) for $\psi_j$ from Eq.(\ref{eqn:sis2}) gives,
\[
 \sum_{r:r|_x=f}\vec{S(\psi_i)}[r] = \sum_{r:r|_x=f}\vec{S(\psi_j)}[r]
\]
which implies $\pi_x(\vec{S(\psi_i)}) =\pi_x(\vec{S(\psi_j)}) $. 

Thus we have a consistent non-trivial super-assignment of norm at most $g$.

\end{proof}

\end{document}